\documentclass[10pt]{extarticle}
\usepackage{graphicx}
\usepackage{amssymb,amsmath}
\def\beq{\begin{eqnarray}}
\def\eeq{\end{eqnarray}}
\usepackage[titletoc,toc,title]{appendix}

\usepackage{amsfonts}

\begin{document}

\title{On the calculation of exact sum rules of rational order for quantum billiards 
	(spectrum with a null eigenvalue)}\author{Paolo Amore \\
\small Facultad de Ciencias, CUICBAS, Universidad de Colima,\\
\small Bernal D\'{i}az del Castillo 340, Colima, Colima, Mexico \\
\small paolo@ucol.mx}

\maketitle

\begin{abstract}
We generalize the calculation of Ref.~\cite{Amore19B} to the case of a spectrum 
containing a zero mode. Using a renormalization procedure, we express the 
sum rules in terms  of suitable traces and show that the final expressions, 
calculated up to second order in perturbation theory agree with the results 
obtained when working directly with the eigenvalues and using Rayleigh-Schrödinger 
perturbation theory.
\end{abstract}

\section{Introduction}
\label{intro}

The purpose of this paper is to generalize the results of Ref.~\cite{Amore19B}, to the case
of a spectrum with a null eigenvalue corresponding, for example, to Neumann or periodic boundary conditions. 
From Refs.~\cite{Amore14,Amore19} where the exact sum rules of integer order have been calculated {\sl to all orders}, 
it is known that one needs to apply a renormalization procedure since the traces containing the contribution of the vanishing eigenvalue are ill--defined. In this case, the expressions for the sum rules differ from the corresponding expressions for a positive definite spectrum, for the presence of extra contributions that appear after renormalization.
Following the procedure of Ref.~\cite{Amore19B}, we extend this renormalization to the case of sum rules of rational 
order, treating the inhomogeneity as a perturbation and working up to second order. 
With this calculation we recover the result of eq.~(9) of Ref.~\cite{Amore12}, which was obtained  working directly
using Rayleigh-Schr\"odinger perturbation theory for the eigenvalues.

The paper is organized as follows: in Section \ref{sec:GF} we introduce the Green's functions of order $1/N$ ($N=2,3,\dots$);
in Section \ref{sec:EN} we add an infinitesimal shift $\gamma>0$ to the Laplacian operator (to obtain a positive definite spectrum)
and obtain the expressions for the energy of the fundamental mode in perturbation theory in the infinitesimal parameter; in 
Section \ref{sec:sumrules} we use the Green's function of rational order to obtain traces of rational order, reproducing 
the result of Ref.~\cite{Amore12}; in Section \ref{sec:num_exp} we verify the general formula for a specific case of a 
linear density with a purely numerical calculation. 
Finally, in Section \ref{sec:concl} we draw our conclusions and discuss possible directions for future work.

\section{Green's functions of order $1/N$}
\label{sec:GF}

In this paper we consider the Helmholtz equation for a heterogeneous system in d dimensions
\begin{equation}
\begin{split}
(-\Delta)_d \Psi_n({\bf x}) = E_n \Sigma(x) \Psi_n({\bf x}) \hspace{.5cm} , \hspace{.5cm} {\bf x} \in  \Omega
\end{split}
\end{equation}
where $\Delta_d\equiv \frac{\partial^2}{\partial x_1^2}+ \dots + \frac{\partial^2}{\partial x_d^2}$ is the Laplacian
operator in d dimensions, $\Sigma({\bf x}) >0$ is a density and $\Omega$ is a d--dimensional region of space.
 The boundary conditions on $\Psi_n({\bf x})$ are  such that the spectrum contains a null eigenvalue, $E_0=0$.

As discussed in Ref.~\cite{Amore14}, in this case the sum rules of integer order, obtained in terms of the traces 
involving the operator $\hat{O} =\frac{1}{\sqrt{\Sigma(x)}} (-\Delta_d) \frac{1}{\sqrt{\Sigma(x)}}$ are ill--defined due to the divergent contribution of the null eigenvalue. The appropriate renormalization procedure to obtain the exact, finite, sum rules, restricted
to non--vanishing eigenvalues has been described in Ref.~\cite{Amore14} for the case of integer order and it will be generalized here
to the case of rational order.

By modifying the operator $\hat{O}$ and performing an infinitesimal shift $\gamma$, 
\begin{equation}
\hat{O}_\gamma \equiv \frac{1}{\sqrt{\Sigma(x)}} (-\Delta +\gamma) \frac{1}{\sqrt{\Sigma(x)}} \ ,
\end{equation}
we manage to work with a positive definite spectrum.

Let us define
\begin{equation}
G_\gamma(x,y) \equiv \sqrt{\Sigma(x)} G_0(x,y) \sqrt{\Sigma(x)}\ ,
\end{equation}
where $\Sigma(x) >0$ is the density and $G_0(x,y)$ is the Green's function of the homogeneous problem ($\Sigma=1$):
\begin{equation}
G_0(x,y) = \sum_n \frac{\psi_n(x) \psi_n^\star(y)}{\epsilon_n+\gamma} =  \frac{1}{\gamma V} + \sum_{q=0} \left(-\gamma\right)^q  G^{(q)}(x,y)\ ,
\end{equation}
where $V \equiv \int_\Omega d^dx$ and 
\begin{equation}
G^{(q)}(x,y) = \sum'_n \frac{\psi_n(x) \psi^\star_n(y)}{\epsilon_n^{q+1}} \ . 
\end{equation}

These functions obey the properties
\begin{equation}
\begin{split}
(-\Delta) G^{(0)}(x,y) &= \delta(x-y)- \frac{1}{V}   \ , \\
(-\Delta) G^{(q)}(x,y) &= G^{(q-1)}(x,y)  \ ,
\end{split}
\end{equation}
and
\begin{equation}
\int G^{(0)}(x,z) G^{(q)}(z,y) = G^{(q+1)}(x,y) \ . 
\end{equation}

It is easy to see that $G_\gamma(x,y)$ is the Green's function associated with the operator $\hat{O}_\gamma$:
\begin{equation}
\begin{split}
\hat{O}_\gamma G_\gamma(x,y) = \frac{1}{\sqrt{\Sigma(x)}} (-\Delta+\gamma) G_0(x,y) \sqrt{\Sigma(y)} = \delta(x-y)
 \ . 
\end{split}
\end{equation}

Now introduce a new function, $\tilde{G}_\gamma^{[1/N]}(x,y)$, satisfying the property
\begin{equation}
\int \tilde{G}_\gamma^{[1/N]}(x,z_1) \tilde{G}_\gamma^{[1/N]}(z_1,z_2) \dots \tilde{G}_\gamma^{[1/N]}(z_N,y) dz_1 dz_2 \dots dz_N  = G_\gamma(x,y)  \ . 
\label{GF_eq_1}
\end{equation}

The property above is fulfilled provided that %$\tilde{G}_\gamma^{[1/N]}(x,y)$ complies with the property
\begin{equation}
\hat{O}_\gamma^{1/N} \tilde{G}_\gamma^{[1/N]}(x,y) = \delta(x-y)  \ .
\end{equation}

Let us now decompose $G_\gamma(x,y)$ in the basis of the unperturbed (homogeneous) problem 
\begin{equation}
G_\gamma(x,y) = \sum_{n,m} Q_{nm} \psi_n(x) \psi^\star_n(y) \ , 
\end{equation}
where
\begin{equation}
\begin{split}
Q_{nm} &= \int \psi_n^\star(x) \sqrt{\Sigma(x)} G_0(x,y) \sqrt{\Sigma(x)} \psi_m(y) dx dy \\
&=  \sum_r \frac{\langle n |\sqrt{\Sigma} | r \rangle \langle r | \sqrt{\Sigma} | m\rangle}{\epsilon_r+\gamma}  \ . 
\end{split}
\end{equation}

Similarly we can decompose $\tilde{G}_\gamma^{[1/N]}$ in this basis as
\begin{equation}
\begin{split}
\tilde{G}(x,y) &= \sum_{n,m} q_{nm}^{[1/N]} \psi_n(x)  \psi_m^\star(y) \ ,
\end{split}
\end{equation}
where
\begin{equation}
q_{nm}^{[1/N]} = \int \int \psi_n^\star(x) \tilde{G}_\gamma^{[1/N]}(x,y) \psi_m(y) dxdy
\end{equation}

Using this expression in the left hand side of eq.~(\ref{GF_eq_1}), we have
\begin{equation}
\begin{split}
{\rm LHS} &= \int \tilde{G}_\gamma^{[1/N]}(x,z_1)\tilde{G}_\gamma^{[1/N]}(z_1,z_2) \dots \tilde{G}_\gamma^{[1/N]}(z_N,y) dz_1 dz_2 \dots dz_N \\
&= \sum_{n,m} \sum_{r_1,\dots,r_N} q_{n r_1}^{[1/N]} \dots q_{r_N m}^{[1/N]} \psi_n(x) \psi_{m}^\star(y)  \ . 
\label{GF_eq_2a}
\end{split}
\end{equation}

Similarly, the right hand side of eq.~(\ref{GF_eq_1}) becomes
\begin{equation}
\begin{split}
{\rm RHS} &= \sum_{n,m} Q_{nm} \psi_n(x) \psi^\star_m(y) \\
&=  \sum_{n,m} \sum_r \frac{\langle n |\sqrt{\Sigma} | r \rangle \langle r | \sqrt{\Sigma} | m\rangle}{\epsilon_r+\gamma} \psi_n(x) \psi^\star_m(y)
\label{GF_eq_2b}
\end{split}
\end{equation}

By equating eqs.~(\ref{GF_eq_2a}) and (\ref{GF_eq_2b}) we finally obtain the matrix equation
\begin{equation}
\begin{split}
\sum_{r_1,\dots,r_N} q_{n r_1}^{[1/N]} \dots q_{r_N m}^{[1/N]}  &= \sum_r \frac{\langle n |\sqrt{\Sigma} | r \rangle \langle r | \sqrt{\Sigma} | m\rangle}{\epsilon_r+\gamma}  \\
&=  \frac{\langle n |\sqrt{\Sigma} | 0 \rangle \langle 0 | \sqrt{\Sigma} | m\rangle}{\gamma}  + \sum'_r \frac{\langle n |\sqrt{\Sigma} | r \rangle \langle r | \sqrt{\Sigma} | m\rangle}{\epsilon_r+\gamma}  \ ,
\end{split}
\label{GF_eq_3}
\end{equation}
where $\sum'_r$ is the sum over all modes with the exclusion of the fundamental mode.

The exact solution of eq.~(\ref{GF_eq_3}), that would provide the exact expression for the Green's function of order $1/N$, cannot be
obtained for an arbitrary density $\Sigma$, and therefore it is convenient to resort to perturbation theory.

In this case we assume a mild inhomogeneity and write
\begin{equation}
\Sigma(x) =  1 + \lambda \sigma(x) \ ,
\end{equation}
with $|\sigma(x)| \ll 1$ for all $x$. $\lambda$ is a power--counting parameter that will be set
to $1$ at the end of the calculation.

Therefore
\begin{equation}
\sqrt{\Sigma(x)} = \sum_{j=0}^\infty \left( \begin{array}{c}
1/2 \\
j \\
\end{array} \right) \lambda^j \sigma(x)^j
\end{equation}

Similarly we can write
\begin{equation}
\begin{split}
q^{[1/N]}_{nm} &= \sum_{j=0}^\infty q_{nm}^{[1/N](j)} \lambda^j  \ ,\\
Q_{nm} &= \sum_{j=0}^\infty Q_{nm}^{(j)} \lambda^j  \ , \\
\label{eq_qQ}
\end{split}
\end{equation}
where
\begin{equation}
\begin{split}
Q_{nm}^{(k)} = \sum_{j=0}^k \left( \begin{array}{c}
1/2 \\
j \\
\end{array} \right) \left( \begin{array}{c}
1/2 \\
k-j \\
\end{array} \right) \left[ \frac{\langle n |\sqrt{\Sigma} | 0 \rangle \langle 0 | \sqrt{\Sigma} | m\rangle}{\gamma}  + \sum'_r \frac{\langle n |\sqrt{\Sigma} | r \rangle \langle r | \sqrt{\Sigma} | m\rangle}{\epsilon_r+\gamma} 
\right]
\end{split}
\end{equation}

Then, by inserting (\ref{eq_qQ}) inside eq.~(\ref{GF_eq_3}) and selecting the term of order $\lambda^k$,
we obtain the matrix equation
\begin{eqnarray}
\sum_{l_{N-1}=0}^k \dots \sum_{l_{1}=0}^{l_2}  \sum_{r_1, \dots, r_N} 
q_{nr_1}^{(l_1)} q_{r_1 r_2}^{(l_2-l_1)} \dots q_{r_N m}^{(k-l_{N-1})} = Q_{nm}^{(k)}
\label{GF_eq_4}
\end{eqnarray}

The solutions of eqs.~(\ref{GF_eq_4}) can be obtained iteratively, starting from the lowest order ($k=0$) and moving to higher orders. To do this it is convenient to introduce the definitions
\begin{equation}
\begin{split}
\Delta_{nm}^{[1/N]}  &\equiv \frac{\left( \frac{1}{\epsilon_n+\gamma} + \frac{1}{\epsilon_m+\gamma} \right)}{\sum_{j=0}^{N-1} \frac{1}{(\epsilon_n+\gamma)^{(N-1-j)/N} (\epsilon_m+\gamma)^{j/N}}}   \ , \\
\eta_{nm}^{[1/N]}  &\equiv \sum_{j=0}^{N-1} \frac{1}{(\epsilon_n+\gamma)^{(N-1-j)/N} (\epsilon_m+\gamma)^{j/N}}  \ ,\\
\xi_{nrm}^{[1/N]}  &\equiv \sum_{j=0}^{N-2} \sum_{l=0}^{N-2-j} \frac{1}{(\epsilon_n+\gamma)^{j/N} (\epsilon_m+\gamma)^{(N-2-j-l)/N} (\epsilon_r+\gamma)^{l/N}} \ .  
\end{split}
\end{equation}

Then, the solutions of eq.~(\ref{GF_eq_4}) up to second order take the 
form~\footnote{The present results can directly be obtained from the corresponding results for the case of a positive definite spectrum of Ref.~\cite{Amore19B} via the substitution $\epsilon_n \rightarrow \epsilon_n + \gamma$.}
\begin{equation}
\begin{split}
q_{nm}^{(0)}  &=   \frac{N}{2} \Delta_{nm}^{[1/N]} \delta_{nm}  \ , \\
q_{nm}^{(1)}  &= \frac{1}{2} \Delta_{nm}^{[1/N]} \langle n |\sigma|m \rangle  \ , \\
q_{nm}^{(2)}  &=  - \frac{1}{8} \Delta_{nm}^{[1/N]} \langle n | \sigma^2 | m \rangle   \\
&+ \frac{1}{4 \eta_{nm}^{[1/N]}} \sum_r \langle n | \sigma | r \rangle \langle r | \sigma | m \rangle \left( \frac{1}{\epsilon_r+\gamma}
-\Delta_{nr}^{[1/N]} \Delta_{rm}^{[1/N]}  \xi_{nrm}^{[1/N]}\right)  \ . 
\end{split}
\end{equation}

\section{Energy of the fundamental mode}
\label{sec:EN}

We consider the Helmholtz equation for the fundamental mode, modified by the presence of the infinitesimal parameter $\gamma$, i.e.,
\begin{equation}
(-\Delta+\gamma)\Psi_0(x) = E_0(\gamma) \Sigma(x) \Psi_0(x) \ ,
\label{eq_Helm}
\end{equation}
where
\begin{equation}
E_0(\gamma) = \sum_{j=1}^\infty \gamma^j E_0^{(j)} \ ,
\end{equation}
and
\begin{equation}
\Psi_0(x) = \sum_{j=0}^{\infty} \gamma^j \Psi_0^{(j)}(x) \ . 
\end{equation}

Notice that, for $j>0$, the corrections to the wave function can be chosen to be orthogonal to the order zero:
\begin{equation}
\int \Psi_0^{(j)}(x) \Psi_0^{(0)}(x) dx = 0
\end{equation}

Substituting the expansions inside eq.~(\ref{eq_Helm}), the corrections to the eigenvalues and eigenfunctions to order $k$ 
are determined by the equation
\begin{equation}
-\Delta \Psi_0^{(k)}(x) + \Psi_0^{(k-1)}(x) = \sum_{j=1}^k E_0^{(j)} \Sigma(x) \Psi_0^{(k-j)}(x) \ . 
\end{equation}
It is easy to see that $\Psi_0^{(0)}(x) = 1/\sqrt{V}$ and $E_0^{(0)}=0$.

Similarly, to first order one obtains 
\begin{equation}
\begin{split}
E_0^{(1)} &= \frac{1}{\langle 0 | \Sigma | 0\rangle} \\
\Psi_0^{(1)}(x) &= \frac{ E_0^{(1)}}{\sqrt{V}}   \int G^{(0)}(x,y) \Sigma(y)  dy  \ . 
\end{split}
\end{equation}

The correction of order $k$ to the eigenvalue and to the eigenfunction are
\begin{subequations}
\begin{align}
E_0^{(k)} &=  - \frac{\sum_{j=1}^{k-1} E_0^{(j)} \langle \psi_0^{(0)} | \Sigma | \psi_0^{(k-j)} \rangle}{\langle \psi_0^{(0)} |  \Sigma | \psi_0^{(0)} \rangle} 
\label{eq_EN_recur}\\
\Psi_0^{(k)}(\Omega) &=  \sum_{j=1}^k E_0^{(j)} \int G^{(0)}(x,y) \Sigma(y) \Psi_0^{(k-j)}(y) dy 
	                 - \int G^{(0)}(x,y) \Psi_0^{(k-1)}(y) dy  \ ,
\label{eq_WF_recur}
\end{align}
\end{subequations}
that can be solved recursively starting from $k=2$. In particular the corrections to $E_0$ up to fourth order read (see Ref.~\cite{Amore19})
\begin{equation}
\begin{split}
E_0^{(2)} &= -\frac{\langle 0 | \Sigma G^{(0)} \Sigma | 0 \rangle}{\langle 0 | \Sigma |0 \rangle^3}  \ ,    \\
E_0^{(3)} &=  \frac{\langle 0 | \Sigma G^{(1)} \Sigma | 0 \rangle}{\langle 0 | \Sigma |0 \rangle^3}  
- \frac{\langle 0 | \Sigma G^{(0)} \Sigma G^{(0)} \Sigma | 0 \rangle}{\langle 0 | \Sigma |0 \rangle^4} 
+ \frac{\langle 0 | \Sigma G^{(0)} \Sigma | 0 \rangle^2}{\langle 0 | \Sigma |0 \rangle^5}   \ ,  \\
E_0^{(4)} &=  -\frac{\langle 0 | \Sigma G^{(2)} \Sigma | 0 \rangle}{\langle 0 | \Sigma |0 \rangle^3} 
+ 2 \frac{\langle 0 | \Sigma G^{(1)} \Sigma G^{(0)} \Sigma | 0 \rangle}{\langle 0 | \Sigma |0 \rangle^4} 
- 4 \frac{\langle 0 | \Sigma G^{(1)} \Sigma | 0 \rangle \langle 0 | \Sigma G^{(0)} \Sigma | 0 \rangle }{\langle 0 | \Sigma |0 \rangle^5} \\
&- 5 \frac{\langle 0 | \Sigma G^{(0)} \Sigma | 0 \rangle^3}{\langle 0 | \Sigma |0 \rangle^7}  +  5 \frac{\langle 0 | \Sigma G^{(0)} \Sigma | 0 \rangle \langle 0 | \Sigma G^{(0)} \Sigma G^{(0)} \Sigma | 0 \rangle}{\langle 0 | \Sigma |0 \rangle^6} \\
&- \frac{\langle 0 | \Sigma G^{(0)} \Sigma G^{(0)} \Sigma G^{(0)} \Sigma| 0 \rangle}{\langle 0 | \Sigma |0 \rangle^5}
  \ .
\end{split}
\end{equation}
Taking into account that $\Sigma = 1 + \sigma$, and expanding up to second order in the density, we
finally obtain
\begin{equation}
\begin{split}
\frac{1}{\left(E_0^{(1)} \gamma +E_0^{(2)} \gamma^2 +\dots \right)^s} & = 
\gamma^{-s} + s \gamma^{-s} \langle 0 |\sigma | 0 \rangle + \frac{1}{2} s (s-1) \gamma^{-s} \langle 0 |\sigma |0 \rangle^2 \\
&+ s \gamma^{1-s} \sum'_n \frac{\langle 0 | \sigma |n \rangle^2}{\epsilon_n} 
- s \gamma^{2-s} \sum'_n \frac{\langle 0 | \sigma |n \rangle^2}{\epsilon_n^2} + \dots
\label{eq_ENs}
\end{split}
\end{equation}

\section{Sum rules of rational order}
\label{sec:sumrules}

The results obtained in the previous two sections allow us to apply the renormalization 
procedure of Refs.~\cite{Amore14,Amore19} to the case of sum rule of rational order.
The sum rule of order $1 +1/N$ can be calculated using the Green's function of order $1$ and $1/N$, as done
in Ref.~\cite{Amore19B} for the case of a positive definite spectrum. 

In this case we obtain
\begin{equation}
\begin{split}
Z\left(1+\frac{1}{N}\right) &= \sum_{n,r} Q_{nr} q_{rn}^{[1/N]}\\
&= \sum_{n,r} Q_{nr}^{(0)} q_{rn}^{[1/N] (0)} \\
&+ \lambda \sum_{n,r} \left[ Q_{nr}^{(0)} q_{rn}^{[1/N] (1)} +  Q_{nr}^{(1)} q_{rn}^{[1/N] (0)}  \right] \\
& + \lambda^2 \sum_{n,r}\left[ Q_{nr}^{(1)} q_{rn}^{[1/N] (1)} + Q_{nr}^{(2)} q_{rn}^{[1/N] (0)} +
Q_{nr}^{(0)} q_{rn}^{[1/N] (2)}  \right] + \dots
\end{split}
\end{equation}

Letting $s=1+1/N$ ($1 < s \leq 3/2$) and substituting the expressions for the $q$ and $Q$ we obtain
\begin{equation}
\begin{split}
Z^{(0)}\left(s\right) &= \frac{1}{\gamma^s} + \sum'_n \frac{1}{(\epsilon_n+\gamma)^s}   \ , \\
Z^{(1)}\left(s\right) &= s \gamma^{-s} \langle 0 |\sigma |0\rangle + \sum'_n \frac{s \langle n | \sigma |n \rangle}{(\epsilon_n +\gamma)^s}   \ , \\
Z^{(2)}\left(s\right) &= \frac{\lambda^2}{2} s (s-1) \gamma^{-s} \langle 0 | \sigma | 0 \rangle^2 + 
\frac{\lambda^2}{2} s (s-1) \sum'_n \frac{\langle n | \sigma | n \rangle^2}{(\epsilon_n+\gamma)^{-s}} \\
&+ s \sum'_n \frac{(\epsilon_n+\gamma)^{1-s}-\gamma^{1-s}}{\epsilon_n} \langle 0 |\sigma | n\rangle^2 \\
&- \frac{s}{2} \sum'_{n\neq m} \frac{(\epsilon_n+\gamma)^{1-s}-(\epsilon_m+\gamma)^{1-s}}{\epsilon_n-\epsilon_m} \langle m|\sigma|n\rangle^2   \ .
\end{split}
\end{equation}

Therefore
\begin{equation}
\begin{split}
Z(s) &= \gamma^{-s} + s \gamma^{-s} \langle 0 |\sigma | 0 \rangle + \frac{1}{2} s (s-1) \gamma^{-s} \langle 0 |\sigma |0 \rangle^2 \\
&+ s \gamma^{1-s} \sum'_n \frac{\langle 0 | \sigma |n \rangle^2}{\epsilon_n} 
- s  \sum'_n \frac{\langle 0 | \sigma |n \rangle^2}{\epsilon_n^s} + \dots \\
&+ \sum'_n \left[\frac{1}{\epsilon_n^s} 
+  \frac{s \langle n | \sigma | n \rangle}{\epsilon_n^s}
+  \frac{s (s-1) \langle n | \sigma | n \rangle^2}{2\epsilon_n^s} \right] \\
&- \frac{s}{2} \sum'_{n \neq m} \frac{\epsilon_n^{1-s}-\epsilon_m^{1-s}}{\epsilon_n - \epsilon_m}
| \langle m |\sigma | n \rangle |^2 + \dots
\end{split}
\end{equation}

Following Ref.~\cite{Amore19} we can define the renormalized sum rule
\begin{eqnarray}
\begin{split}
\tilde{Z}(s) &\equiv \lim_{\gamma \rightarrow 0} \left(Z(s) - \frac{1}{\left(E_0^{(1)} \gamma +E_0^{(2)} \gamma^2 +\dots \right)^s}  \right) \\
&= \sum'_n \left[\frac{1}{\epsilon_n^s} 
+  \frac{s \langle n | \sigma | n \rangle}{\epsilon_n^s}
+  \frac{s (s-1) \langle n | \sigma | n \rangle^2}{2\epsilon_n^s} \right] \\
&- \frac{s}{2} \sum'_{n \neq m} \frac{\epsilon_n^{1-s}-\epsilon_m^{1-s}}{\epsilon_n - \epsilon_m}
| \langle m |\sigma | n \rangle |^2 - s \sum'_n  \frac{|\langle 0| \sigma|n\rangle|^2}{\epsilon_n^s} + \dots   \ ,
\end{split}
\end{eqnarray}
where the last term in this expression is a {\sl finite}  contribution stemming from the zero mode.
This result can be condensed into the formula
\begin{equation}
\begin{split}
\tilde{Z}(s)  &= \sum'_n \left[\frac{1}{\epsilon_n^s} 
+  \frac{s \langle n | \sigma | n \rangle}{\epsilon_n^s}
+  \frac{s (s-1) \langle n | \sigma | n \rangle^2}{2\epsilon_n^s} \right] 
- \frac{s}{2} \sum_{n \neq m} \frac{\epsilon_n^{1-s}-\epsilon_m^{1-s}}{\epsilon_n - \epsilon_m}
| \langle m |\sigma | n \rangle |^2 + \dots    
\end{split}
\label{eq_ZPT}
\end{equation}
This is precisely the eq.(9) of Ref.~\cite{Amore12}, obtained using Rayleigh-Schr\"odinger perturbation theory.

%Notice that the same result is also obtained when considering the sum rule of order $1/2+1/N$ (this sum rule however converges only in one dimension).

%\subsection{$Z(1/2+1/N)$}

\section{A numerical experiment}
\label{sec:num_exp}

We consider a heterogeneous string of unit length ($|x| \leq 1/2$) and with density 
\begin{equation}
\Sigma(x) = 1 + \kappa x  \ , 
\end{equation}
and assume Neumann boundary conditions. The condition $\Sigma(x) >0$ for $x \in (-1/2,1/2)$ implies $|\kappa|< 2$.
The general expression for the exact sum rule of order $1$ for a heterogeneous string has been derived 
in Ref.~\cite{Amore14} and it reads
\begin{equation}
\begin{split}
Z(1) &= \int_{-1/2}^{1/2} \left( \frac{1}{12} + x^2 \right) dx
- \frac{\int_{-1/2}^{1/2} dx \int_{-1/2}^{1/2} dy \Sigma(x) G^{(0)}(x,y) \Sigma(y)}{\int_{-1/2}^{1/2} \Sigma(x) dx }    \ , 
\label{eq_Z1}
\end{split}
\end{equation}
where $G^{(0)}(x,y)$ is the regularized Green's function for Neumann bc~\cite{Amore13A}
\begin{equation}
G^{(0)}(x,y) = \frac{1-6 |x-y| + 6 (x^2+y^2)}{12} \  .
\end{equation}
Notice that the second term in eq.(\ref{eq_Z1}) originates from the renormalization of the trace. 

In the present case one obtains 
\begin{equation}
Z(1) = \frac{1}{6} - \frac{\kappa^2}{120} \  .
\end{equation}
This result is exact {\sl to all orders} in the density.

The calculation can be also carried out using eq.~(\ref{eq_ZPT}), obtaining 
\begin{equation}
Z(1) = \frac{1}{6} - \frac{8 \kappa ^2}{\pi^6} \sum_{n=1}^\infty \frac{1}{(1-2 n)^6} =  \frac{1}{6} - \frac{\kappa^2}{120}
\end{equation}
which confirms the exact result.

For the case of sum rules of non--integer order, results that are exact to all orders are not available and therefore one needs to
rely on numerical results to assess the perturbative formula in eq.~(\ref{eq_ZPT}). Accurate estimates for the lowest
eigenvalues of the string can be obtained applying the Rayleigh-Ritz method; on the other hand, the highest part of the spectrum of a heterogeneous string can also be estimated precisely using the approach of Ref.~\cite{Amore11}.

In particular, the leading asymptotic behavior of the eigenvalues of the string as $n \rightarrow \infty$ is
\begin{equation}
E_n^{(asym)} = \frac{18 \pi^2 \kappa^2 n^2}{\left((2-\kappa)^{3/2}-(2+\kappa)^{3/2}\right)^2} + \dots
\end{equation}

The sum rule can be approximated as
\begin{equation}
Z^{(num)}(1) = \sum_{n=1}^{n_{max}} \frac{1}{\left( E_n^{(RR)} \right)^s} + \sum_{n=n_{max}+1}^\infty \frac{1}{\left( E_n^{(asym)} \right)^s} \  ,
\end{equation}
where $E_n^{(RR)}$ are the numerical eigenvalues obtained with the Rayleigh-Ritz method and $n_{max}$ is a cutoff (in our case
$n_{max}=200$).

The sum rule of order $3/2$ calculated perturbatively up to second order is
\begin{equation}
\begin{split}
Z(3/2) &= \frac{\zeta(3)}{\pi^3} \\
 &+ \left[ -\frac{381 \zeta (7)}{32 \pi ^7} + \sum_{n=1}^\infty \sum_{m=1}^\infty \frac{12 \left(4 m^2+(1-2 n)^2\right)^2}{\pi^7 m (2 n-1) (2 m-2 n+1)^4 (2 m+2 n-1)^5} \right] \kappa ^2  \\
&\approx \frac{\zeta(3)}{\pi^3} + \left[ -\frac{381 \zeta (7)}{32 \pi ^7} + 0.000539831 \right] \kappa ^2 \\
&\approx 0.03876817960-0.00343517 \kappa ^2  \  .
\end{split}
\end{equation}

We have used the numerical eigenvalues calculated using the Rayleigh-Ritz method with $2001$ modes 
to obtain numerical  approximations of $Z(3/2)$ at $20$ values of $\kappa$ ($\kappa=1/100, 2/100, \dots, 1/5$). 
These results have then been fitted with polynomial of order four in $\kappa$, thus obtaining the estimate
\begin{equation}
\begin{split}
Z^{({\rm fit})}(3/2) & \approx  0.0387682 + 3.3 \times 10^{-10} \kappa -0.00343517 \kappa ^2 \\
&+ 1.76 \times 10^{-8} \kappa^3 + 0.0000971414 \kappa ^4  \  ,
\end{split}
\end{equation}
that is remarkably close to the exact result.

\section{Conclusions}
\label{sec:concl}

We have calculated the sum rules of rational order for the eigenvalues of the Helmholtz equation in $d$ dimensions in presence of a 
heterogeneous medium and for boundary conditions allowing for a null eigenvalue. From our previous works, refs.~\cite{Amore14, Amore19}, 
it is known that the sum rules at integer exponents can be cast in terms of traces involving suitable Green's functions by following the renormalization procedure  
originally introduced in Ref.~\cite{Amore14}. In this renormalization it is seen that extra terms appear compared with the analogous case for boundary conditions
not allowing a null eigenvalue. The calculation of a sum rule in terms of a trace, which in our view is more rigorous than the approach of Ref.~\cite{Amore12}, 
involves Green's functions that are decoupled from the zero mode. The renormalization generates extra contributions which couple to the zero mode.
By extending the calculation of Ref.~\cite{Amore19B} to the case of a spectrum with a null eigenvalue and following the
renormalization of Ref.~\cite{Amore14}, we recover the formula (9) of Ref.~\cite{Amore12}, was obtained using the Rayleigh-Schr\"odinger perturbation theory.

\section*{Acknowledgements}
I am grateful to Prof. A.J. Stuart for reading this manuscript and for useful suggestions.
This research was supported by the Sistema Nacional de Investigadores (M\'exico).

%%%%%%%%%%%%%%%%%%%%%%%%%%%%%%%%%%%%%%%%%%%%%%%%%%%%%%%%%%%%%%%%%%%%%%%%%%%%%%%%%%%%%%%%%%%%%%%%%%%%%%%%%%%%%%%%%%%%%%%%%%%%%%%%%%%%%%%


\begin{thebibliography}{Bibliography}
	
\bibitem{Amore19B} Amore, Paolo, "On the calculation of exact sum rules of rational order for quantum billiards" (2019)
\bibitem{Amore14} Amore, Paolo. "Exact sum rules for inhomogeneous systems containing a zero mode." Annals of Physics 349 (2014): 253-267.
\bibitem{Amore19} Amore, Paolo, "Exact sum rules for heterogeneous spherical drums", math-ph, arXiv:1907.10034 (2019)
\bibitem{Amore12} Amore, Paolo. "A perturbative approach to the spectral zeta functions of strings, drums, and quantum billiards." 
Journal of Mathematical Physics 53.12 (2012): 123519.	
\bibitem{Amore11} Amore, Paolo, "The string of variable density: Further results", Annals of Physics {\bf 326}, 2315-2355 (2011)
\bibitem{Amore13A} Amore, Paolo. "Exact sum rules for inhomogeneous strings." Annals of Physics 338 (2013): 341-360.


%\bibitem{Amore13B} Amore, Paolo. "Exact sum rules for inhomogeneous drums." Annals of Physics 336 (2013): 223-244.
%\bibitem{Amore18} Amore, Paolo, "Exact sum rules for quantum billiards of arbitrary shape", Annals of Physics 388 (2018): 12-24.
\end{thebibliography}
\end{document}